\documentclass[aps,prl,twocolumn,notitlepage,superscriptaddress]{revtex4-1}
\usepackage{amsmath}
\usepackage{amssymb}
\usepackage{amsfonts}
\usepackage{color}
\usepackage{dcolumn}
\usepackage[cp1251]{inputenc}
\usepackage[english]{babel}
\usepackage{epsfig}
\usepackage{array}
\usepackage{dsfont}
\usepackage{comment}
\usepackage[all]{xy}
\usepackage{amsmath} 
\usepackage{amsthm} 
\usepackage{amssymb}	
\usepackage{graphicx} 

\renewcommand{\v}[1]{\ensuremath{\mathbf{#1}}} 
 
\def\tr{\textup{Tr}\,}

\def\schrodinger{Schr\"{o}dinger }
\def\one{\mathds{1}} 

\newcommand{\mean}[1]{\langle #1 \rangle}           

\newcommand{\ket}[1]{|#1\rangle}                    
\newcommand{\bra}[1]{\langle #1|}                   

\newcommand{\braket}[2]{\left< #1 \vphantom{#2} \right|\left.\!\! #2 \vphantom{#1} \right>} 

\newcommand{\oprod}[2]{\ket{#1}\bra{#2}}
\newcommand{\proj}[1]{\oprod{#1}{#1}}

\includecomment{details}

\begin{document}
\title{Device-independent quantum key distribution with generalized two-mode \schrodinger cat states}
\author{Curtis J. Broadbent}
\affiliation{Rochester Theory Center, University of Rochester, Rochester, NY 14627}
\affiliation{Center for Coherence and Quantum Optics, University of Rochester, Rochester, NY 14627}
\affiliation{Department of Physics and Astronomy, University of Rochester, Rochester, NY 14627}
\author{Kevin Marshall}
\affiliation{Department of Physics, University of Toronto, Toronto, ON M5S 1A7, Canada}
\author{Christian Weedbrook}
\affiliation{QKD Corp., 60 St.~George St., Toronto, M5S 1A7, Canada}
\author{John C.  Howell}
\affiliation{Center for Coherence and Quantum Optics, University of Rochester, Rochester, NY 14627}
\affiliation{Department of Physics and Astronomy, University of Rochester, Rochester, NY 14627}

\date{\today}
\begin{abstract}
We show how weak non-linearities can be used in a device-independent quantum key distribution (QKD) protocol using generalized two-mode \schrodinger cat states. The QKD protocol is therefore shown to be secure against collective attacks and for some coherent attacks. We derive analytical formulas for the optimal values of the Bell parameter, the quantum bit error rate, and the device-independent secret key rate in the noiseless lossy bosonic channel. Additionally, we give the filters and measurements which achieve these optimal values. We find that over any distance in this channel the quantum bit error rate is identically zero, in principle, and the states in the protocol are always able to violate a Bell inequality. The protocol is found to be superior in some regimes to a device-independent QKD protocol based on polarization entangled states in a depolarizing channel. Finally, we propose an implementation for the optimal filters and measurements.
\end{abstract}

\maketitle
The last two decades have seen a rise in the number and quality of quantum key distribution (QKD) protocols \cite{Scarani2009,Weedbrook2012,Lo2014}. Some of these have been developed into successful commercial products currently deployed in telecommunications \cite{idquantique2014,quintessence2014}. These systems are designed on the principle of provably secure communication, in which, under certain assumptions, the security is guaranteed by the laws of physics, not the assumed difficulty of performing certain mathematical operations as in classical cryptography protocols. Unfortunately, practical implementations of QKD protocols have in many cases fallen short of their desired goal; due to rate ceilings, current QKD systems are used only to generate keys for use with standard cryptographic protocols. Additionally, in recent years both research and commercially developed QKD systems have been successfully hacked using side-channel information \cite{Makarov2006,Makarov2008,Lamas-Linares2007,Zhao2008,Xu2010,Lydersen2010,Lydersen2010a,Yuan2010,Gerhardt2011,Weier2011,Jain2011,Sun2011,Tang2013,Jouguet2013,Huang2013,Huang2014}.  

In response to these limitations, there has been an effort to develop QKD protocols which are immune to the practical limitations of the devices in which they are implemented. These protocols are called device-independent QKD (diQKD) protocols and are based on violation of Bell and EPR-steering inequalities \cite{Mayers2004,Acin2007,Pironio2009,Branciard2012,Marshall2014}. If a particular physical implementation of the device-independent QKD protocol is able to violate a Bell inequality, then the resulting key can be considered to be secure, regardless of the details of the physical implementation. Device-independent QKD protocols have been shown to be secure under collective attacks and in some instances are secure under coherent attacks \cite{Acin2007,Pironio2009,McKague2009,Masanes2011}.   

Though device-independence provides a way around the security limitations of previous QKD protocols, it further restricts the secret key generation rate which may be obtained. As a result, there is a growing interest in trying to implement diQKD in diverse systems in an attempt to increase the secret key generation rate. With that in mind, in this manuscript we present an alternative implementation of diQKD which makes use of highly non-Gaussian states: phase-entangled coherent states. We show here that in certain bosonic channels, phase-entangled coherent states have secret key rates competitive with state-of-the-art diQKD systems, including diQKD systems based on discrete variables. Beyond being competitive with state of the art systems, we also show that phase-entangled coherent states allow for a high degree of flexibility in the deployment of the QKD protocol. This could be helpful in situations where the properties of the channel, in particular, the total transmission rate, are variable over time. Additionally, though the phase-entangled coherent state is non-Gaussian, it is relatively easy to generate, requiring only a single photon source and a relatively weak Kerr non-linearity \cite{Kirby2014}. 

Device-independent quantum key distribution proceeds in the following manner. Two distant parties, Alice and Bob, receive entangled pairs from a distant third party, who is possibly under the control of an eavesdropper, Eve. We assume that the measurement device may not be trustworthy having possibly been manufactured by Eve. Alice and Bob are able to set their devices to measure the operators $\{A_0,A_1,A_2\}$ and $\{B_1,B_2\}$ respectively; they record outcomes $\{a_i,b_j\}$ which take values in $\{-1,1\}$. We assume that Alice and Bob are in secure locations; they are able to confirm that their devices do not transmit signals of any kind to Eve. Alice and Bob select their measurement settings at random for each photon they measure. After all measurements have been performed, Alice and Bob communicate the bases in which measurements where made. Measurements performed in the bases $\{A_0,B_1\}$ are used to generate a secret key and determine the quantum bit error rate (QBER) $Q=\textup{prob}(a_0\neq b_1)$. Measurements performed in the basis $\{A_i,B_j\}$ $(i,j\in\{1,2\})$ are used to determine whether a Bell inequality is violated. Measurements performed in the basis $\{A_0,B_2\}$ are assumed to be uncorrelated and are discarded.

We consider the phase-entangled coherent state which is created by the entangling interferometer described in \cite{Kirby2013}, 
\begin{align}
  \ket{\psi}=(\ket{\alpha_+}\ket{\alpha_-}-\ket{\alpha_-}\ket{\alpha_+})/N,
\end{align}
where $\ket{\alpha_\pm}$ are coherent states with oppositely rotated phases $\ket{\alpha_\pm}=\ket{\alpha e^{\pm i\phi}}$ with $\alpha\geq0$ where $N=\sqrt{2(1-\gamma^2)}$ and $\gamma=|\braket{\alpha_+}{\alpha_-}|$. The phase-entangled coherent states are equivalent to displaced \schrodinger cat states, 
\begin{equation}
  \begin{aligned}
  \ket{\psi}=&\left(\mathcal{D}(\alpha \cos\phi)\otimes\mathcal{D}(\alpha \cos\phi)\right)\times\\
 &\quad \left(\ket{\beta}\ket{-\beta}-\ket{-\beta}\ket{\beta}\right)/N,
\end{aligned}
\end{equation}
where $\beta=i\alpha\sin\phi$. Additionally we note that antisymmetric states of this type have a concurrence of 1, are relatively robust to photon loss and are able to violate a Bell inequality after a distance 400 km using state discrimination techniques assumming loss of $0.15$dB/km \cite{Kirby2014}. In the following, we show that in principle, with optimal measurements, the phase-entangled coherent states can give rise to Bell inequality violation at any distance. It is this property we leverage for our device-independent QKD protocol.

Once generated, the phase-entangled coherent states pass through an optical fiber to Alice and Bob. We model the loss incurred in the fiber as beam-splitter loss with transmission and reflection coefficients $t$ and $r$, respectively, where $|t|^2+|r|^2=1$, and where the other input mode of the beam-splitter is in the vacuum state. The resulting state is given by
\begin{align}
  \ket{\psi'}&=\bigg(\ket{t\alpha_+}\ket{t\alpha_-}\ket{r\alpha_+}_L\ket{r\alpha_-}_L\nonumber\\&\quad\quad\quad\quad\quad-\ket{t\alpha_-}\ket{t\alpha_+}\ket{r\alpha_-}_L\ket{r\alpha_+}_L\bigg)/N\label{afterfiber},
\end{align}
where we have assumed for simplicity that the loss coefficients for the fiber to Alice are equal to those for the fiber to Bob. The subscript $L$ in \eqref{afterfiber} refers to the lossy modes of the optical fiber, where we establish the convention that in products of the form $\ket{\cdot}_L\ket{\cdot}_L$ the first ket always refers to the lossy modes of Alice's fiber and the second to Bob's, and single kets of the form $\ket{\cdot}_L$ refer generally to the joint space of Alice and Bob's lossy modes. 

We define the orthonormal states following the work of \cite{Lastra2010}, which will be of use in tracing out the lossy modes and quantifying the entanglement of the remaining state, 
\begin{subequations}
\begin{align}
  \ket{\pm}&=\frac{1}{N_\pm}\bigg(e^{i\delta_t/2}\ket{t\alpha_+}\pm e^{-i\delta_t/2}\ket{t\alpha_-}\bigg),\\
  \ket{\pm}_L&=\frac{1}{M_\pm}\bigg(\ket{r\alpha_+}_L\ket{r\alpha_-}_L\pm\ket{r\alpha_-}_L\ket{r\alpha_+}_L\bigg),
\end{align}\label{orthogonalbasis}
\end{subequations}
where $N_\pm=\sqrt{2(1\pm\gamma_t)}$ and $M_\pm=\sqrt{2(1\pm\gamma_r^2)}$, with $\gamma_q=|\braket{q\alpha_+}{q\alpha_-}|=\gamma^{|q|^2}$ and $\delta_t=\arg\braket{t\alpha_+}{t\alpha_-}$. It will also be helpful to write out the inverse relationships,
\begin{subequations}
    \begin{align}
  \ket{t\alpha_\pm}&=\frac{e^{\mp i\delta_t/2}}{2}\bigg(N_+\ket{+}\pm N_-\ket{-}\bigg),\\
  \ket{r\alpha_\pm}_L\ket{r\alpha_\mp}_L&=\frac{1}{2}\bigg(M_+\ket{+}_L\pm M_-\ket{-}_L\bigg).
\end{align}\label{inversebasis}
\end{subequations}
Substituting \eqref{inversebasis} into \eqref{afterfiber} and tracing over the lossy modes, we arrive at the following mixed state written in the basis of $\{\ket{++},\ket{+-},\ket{-+},\ket{--}\}$
\begin{align}
  \rho&=\frac{M_-^2}{16N^2}(N_+^4+N_-^4)\proj{\Psi}+\frac{M_+^2}{16N^2}(2N_+^2N_-^2)\proj{\Phi},
\end{align}
where $\ket{\Psi}=(N_+^2\ket{++}-N_-^2\ket{--})/\sqrt{N_+^4+N_-^4}$ and $\ket{\Phi}=(\ket{+-}-\ket{-+})/\sqrt{2}$.

We use this form of $\rho$ to determine the optimal filtering operations for single-copy entanglement distillation. The optimal filters $M_A\otimes M_B$ are those which result in a Bell-diagonal state, have maximum probability of success, and satisfy the requirements of being a valid filtering operation \cite{Verstraete2001}. A Bell-diagonal state is one which can be expressed as a mixture of Bell states. The probability of success of the filtering operation is given by 
\begin{align}
  p=\tr(M_A\otimes M_B)\rho(M_A\otimes M_B)^\dagger.
\end{align}
Finally, a valid filtering operation $M_i$ has the property that $(\one-M_i^\dagger M_i)$ is a positive operator. We introduce the state $\varrho$ to simplify the algebra,
\begin{align}
  \varrho=\left(
  \begin{matrix}
    a & 0 & 0 &-\sqrt{ad}\\
    0 & b & -b & 0\\
    0 & -b & b & 0\\
    -\sqrt{ad} & 0 & 0 & d
  \end{matrix}
  \right),
\end{align}
where the quantities in $\varrho$ are defined by the relation $\varrho=\rho$, i.e. $a=M_-^2N_+^4/(16N^2)$, $b=M_+^2N_+^2N_-^2/(16N^2)$, and $d=M_-^2N_-^4/(16N^2)$. Consequently, we note that $a\geq d$ since $N_+\geq N_-$. Using the method described by \citet{Verstraete2001} it can be shown that the optimal filtering operations are given by
\begin{align}
  M_A=M_B=\left(
  \begin{matrix}
    (d/a)^{1/4} & 0\\
    0 & 1\\
  \end{matrix}
\right),
\end{align}
also written as $M_A=M_B=(d/a)^{1/4}\proj{+}+\proj{-}$. The state after the filtering operation is given by 
\begin{align}
  \varrho'=\frac{1}{2(b+\sqrt{ad})}\left(
  \begin{matrix}
    \sqrt{ad} & 0 & 0 &-\sqrt{ad}\\
    0 & b & -b & 0\\
    0 & -b & b & 0\\
    -\sqrt{ad} & 0 & 0 & \sqrt{ad}\\
  \end{matrix}
  \right),\label{varrhoprime}
\end{align}
which occurs with a probability of success of $p=2\sqrt{d/a}(b+\sqrt{ad})$. Since $a$, $b$, and $d$ are all positive, we see that the filtering operation always has a non-zero probability of success.

Following the work of \citet{Horodecki1995}, we can use the correlation matrix to determine the maximum possible Bell violation after the filtering operations. We briefly review how this is done. The correlation matrix $C$ is defined to be $C_{ij}=\tr(\sigma_i\otimes\sigma_j)\varrho'$, where $\sigma_i$ and $\sigma_j$ are the standard Pauli matrices in the basis $\{\ket{+},\ket{-}\}$, with $i,j\in\{x,y,z\}$. A dichotomic measurement operator $A$ with eigenvalues of $\pm1$ can be represented by the unit vector $\mathbf{a}=(a_x,a_y,a_z)$ where $A=a_x\sigma_x+a_y\sigma_y+a_z\sigma_z$. It is straightforward to show that a joint dichotomic measurement $\mean{A\otimes B}$ can be calculated using the correlation matrix as $\mean{A\otimes B}=\mathbf{a}.C.\mathbf{b}$, where we define $\mathbf{b}$ for operator $B$ similar to $\mathbf{a}$. Consequently, the Bell parameter is given by
\begin{align}
  \mean{S}=(\mathbf{a}_1+\mathbf{a}_2).C.\mathbf{b}+(\mathbf{a}_1-\mathbf{a}_2).C.\mathbf{b}_2,\label{vectorform}
\end{align}
where $S=A_1\otimes B_1+A_2\otimes B_1+A_1\otimes B_2-A_2\otimes B_2$. The form of the Bell parameter given in \eqref{vectorform} lends itself easily to constrained maximization techniques as shown by \citet{Horodecki1995}. The maximum value of the Bell parameter is given by $S_{max}=2\sqrt{s_1^2+s_2^2}$ where $s_1$ and $s_2$ are the largest two singular values of the correlation matrix. The left- and right-singular vectors $\mathbf{l}_i$ and $\mathbf{r}_i$ of the correlation matrix can be used to determine a set of measurements which acheives the maximum value of the Bell parameter. Specifically, 
\begin{equation}
\begin{aligned}
  \mathbf{a}_1&=\cos\varphi\,\mathbf{l}_1+\sin\varphi\,\mathbf{l}_2,&\mathbf{b}_1&=\mathbf{r}_1,\\
  \mathbf{a}_2&=\cos\varphi\,\mathbf{l}_1-\sin\varphi\,\mathbf{l}_2,&\mathbf{b}_2&=\mathbf{r}_2,
 \end{aligned}\label{measurements}
 \end{equation}
 where $\cos\varphi=s_1/\sqrt{s_1^2+s_2^2}$ with $(0\leq\varphi\leq\pi/4)$. Mapping the vectors in \eqref{measurements} back to operators in the two-qubit Hilbert spaces yields the appropriate measurements for achieving the maximum value of the Bell parameter experimentally. 

In our specific case, i.e. for the state $\varrho'$, the correlation matrix is diagonal with values $C_{xx}=-1$ and $C_{yy}=C_{zz}=-(b-\sqrt{ad})/(b+\sqrt{ad})$. Consequently, the maximum value of the Bell parameter is 
\begin{align}
  S_{\textup{max}}=2\sqrt{1+\left(\frac{b-\sqrt{ad}}{b+\sqrt{ad}}\right)^2}.\label{maxvalue}
\end{align}
Except for the case where $\sqrt{ad}=b$, the state $\varrho'$ is always able to violate a Bell inequality. A set of measurements which give rise to the maximum value of the Bell parameter are
\begin{equation}
\begin{aligned}
 A_1&=\cos\varphi\,\sigma_x+\sin\varphi\,\sigma_y,&B_1&=-\sigma_x,\\
 A_2&=\cos\varphi\,\sigma_x-\sin\varphi\,\sigma_y,&B_2&=-\sigma_y,
 \end{aligned}\label{measurements}
 \end{equation}
 where $\cos\varphi=2/S_{\textup{max}}$ with $(0\leq\varphi\leq\pi/4)$.

 We now express the probability of success and the maximum value of the Bell parameter in terms of the initial overlap $\gamma$ and the total transmission probability $T=|t|^2$,
 \begin{align}
   p&=\frac{(1-\gamma^T)^2}{1-\gamma^2},\\
   S_{\textup{max}}&=2\sqrt{1+\gamma^{4(1-T)}}.\label{longdistbell}
 \end{align}
We see in \eqref{longdistbell} that for a pure-loss channel, optimally filtered phase-entangled coherent states are able to violate a Bell inequality at any distance. Finally, as can be deduced from the value of $C_{xx}$, measurements made in the $\sigma_x\otimes\sigma_x$ basis are always perfectly anti-correlated regardless of the value of $\gamma$ or $T$. If $A_0=\sigma_x$ then, in principle, the QBER for the phase entangled coherent states will be exactly zero, $Q_{\textup{pecs}}=0$. These two remarkable results, Bell inequality violation as well as zero QBER at any distance are a result of the noiseless nature of the pure-loss channel considered here, and are not expected to extend to channel models which include noise. 

\begin{figure}
\begin{centering}
\includegraphics[width=\columnwidth]{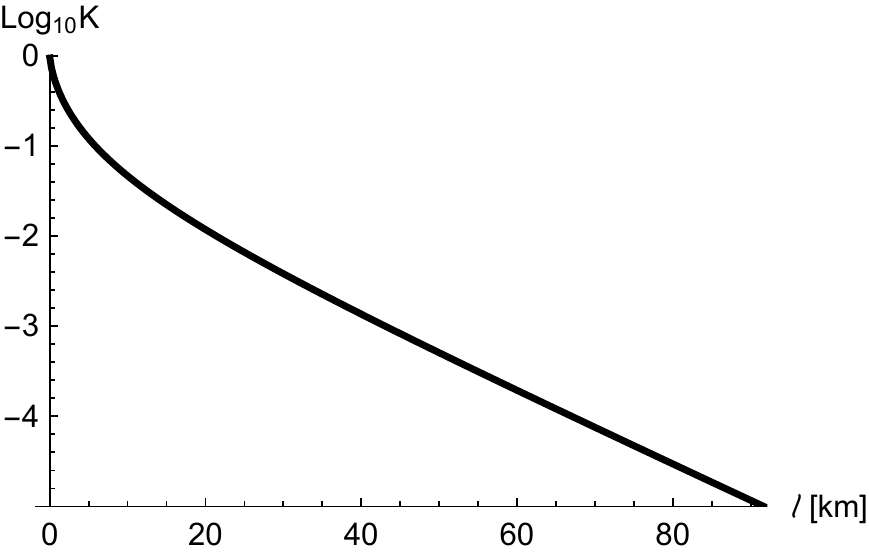}
\caption{The secret key fraction $K$ as a function of the distance between Alice and Bob assuming a loss coefficient of $0.2$ dB/km and the optimal $\gamma$ for the specified distance. This plot gives the raw fraction; it does not include finite detection efficiencies, dark counts, thermal noise, or the relative sampling frequency of the $\{A_0,B_1\}$ measurement.}\label{secretK}
\end{centering}
\end{figure}

We can use the probability of success, the maximum value of the Bell parameter, and the QBER to determine a secret key rate for a device-independent implementation of the protocol. This is done by noting that the Holevo information between Eve and Bob can be bounded using the Bell violation,
\begin{align}
  \chi(B_1,E)\leq h\bigg(\frac{1+\sqrt{(S/2)^2-1}}{2}\bigg),\label{holevo}
\end{align}
where the Holevo information is equal to the quantum mutual information of the joint state shared by Bob and Eve after Bob's measurements have been performed \cite{Acin2007}. In \eqref{holevo}, $h$ is the binary Shannon entropy. Noting that the mutual information between $A_0$ and $B_1$ is 1 and using the Holevo information in \eqref{holevo}, we find that the Devetak-Winter rate for the optimally filtered state is given by 
\begin{align}
  r_{DW}=1-h\bigg(\frac{1+\gamma^{2(1-T)}}{2}\bigg).
\end{align}
Consequently, the raw secret key fraction $K$, under collective attacks and with one-way classical post-processing from Bob to Alice is bounded by the probability of successful filtering times the Devetak-Winter rate, 
\begin{align}
  K\geq p\,r_{DW} = \frac{(1-\gamma^T)^2}{1-\gamma^2}\bigg(1-h(\frac{1+\gamma^{2(1-T)}}{2})\bigg).
\end{align} 
Since $\gamma$ is a free parameter to be set by Alice and Bob given knowledge of the total transmission T, we may optimize the secret key fraction as a function of $\gamma$. We plot the resulting secret key fraction in Fig.~\ref{secretK}. The total secret key rate is just the product of the secret key fraction, the $\{A_0,B_1\}$ sampling probability, and the raw repetition rate of the source.

For large distances, to lowest order in the total transmission $T$, the optimal secret key fraction scales as $T^2$. The scaling coefficient is given by 
\begin{align}
  \alpha = \frac{\log^2\gamma}{1-\gamma^2}\bigg(1-h(\frac{1+\gamma^2}{2})\bigg),
\end{align}
which has a maximum of approximately $4.6\%$ when $\gamma\simeq0.74$. By comparison, due solely to loss, the secret key fraction for a biphoton entangled source scales no better than $T^2$ irrespective of the QKD protocol. This indicates that compared against an error-free biphoton-based QKD protocol, our protocol can never have a secret key rate worse than $4.6\%$ of that which can be achieved in a biphoton-based protocol. In the presence of noise, however, our protocol can drastically outperform some biphoton-based diQKD protocols, as we now discuss.

\begin{figure}
\begin{centering}
\includegraphics[width=\columnwidth]{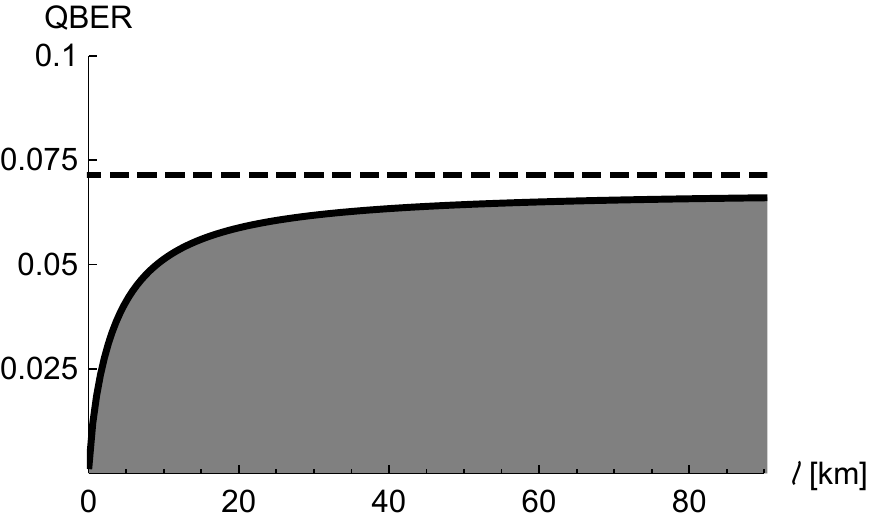}
\caption{Plot of the critical QBER $Q_{\textup{crit.}}$ as a function of distance, see text for details. The biphoton approach is superior in the gray region, corresponding to $Q<Q_{\textup{crit.}}$. The phase-entangled coherent state approach is superior in the white region, corresponding to $Q>Q_{\textup{crit.}}$. The discrete variable diQKD approach using biphotons cannot be used when the QBER exceeds the dashed black line.}\label{equivQBER}
\end{centering}
\end{figure}

We make a comparison to a diQKD protocol based on polarization entangled states subject to depolarizing noise \cite{Pironio2009}. In this case, depolarizing noise increases the probability of error and decreases the Bell violation. The resulting secret key fraction has been shown to be given by
\begin{align}
  K_{\textup{biphoton}}=T^2\bigg(1-h(Q)-h\bigg[\frac{1+\sqrt{(S/2)^2-1}}{2}\bigg]\bigg),
\end{align}
where $S=2\sqrt{2}(1-2Q)$ \cite{Pironio2009}. To permit a comparison to known results, we would like to determine values of the QBER and the propagation distance for which the phase-entangled coherent state based protocol gives an improved secret key fraction as compared to biphoton based QKD protocols, i.e. $K\geq K_{\textup{biphoton}}$. We calculate numerically the critical QBER $Q_{\textup{crit.}}$ such that $K_{\textup{biphoton}}=K$ at a given distance. When the QBER exceeds $Q_{\textup{crit.}}$, the protocol based on the phase-entangled coherent state will have a higher secret key rate than the biphoton-based protocol, i.e. for $Q>Q_{\textup{crit.}}$ we find that $K>K_{\textup{biphoton}}$. We plot this critical QBER as a function of distance in Fig.~\ref{equivQBER}. We find that our approach using \schrodinger cat states is superior whenever the QBER for the depolarizing channel exceeds the black line in Fig.~\ref{equivQBER}. For example, we find that at a distance of 11 km, if the quantum bit error rate is $6.6\%$, our protocol has a secret key rate roughly twice that of the biphoton-based protocol. It can be seen that at large distances, our approach is superior when the QBER of the depolarizing channel exceeds $6.7\%$. Further, it is clear that there are QBERs for which the depolarizing channel cannot be used at all ($Q\geq0.715$), whereas the phase-entangled coherent state approach is viable over any distance. This analysis, which is valid only for a pure-loss channel for the phase-entangled diQKD protocol, suggests that there may be physical channels in which the coherent state approach is superior to alternative protocols, even in the presence of noise.  

Examination of Fig.~\ref{secretK} and Fig.~\ref{equivQBER} reveals another advantage of the \schrodinger cat implementation of diQKD; by tuning $\gamma$ to the transmission distance, the secret key fraction can be improved dramatically for short distances relative to the long term scaling. This sort of dynamical control is not typically present in standard implementations of diQKD in discrete systems. 

We now discuss the feasibility of performing the measurements required to implement the optimal filtering and to violate the Bell inequality. The primary requirement is the ability to make projective measurements in a superposition of the $\ket{+}$ and $\ket{-}$ states. Techniques for making such a measurement are the subject of ongoing investigations, but recent theoretical results for Gaussian states as well as advances in detectors suggest that these types of measurements are becoming easier to implement experimentally. 

We briefly describe one possible approach which could be used to perform the required measurements using presently available equipment. The primary difficulty is to measure in an arbitrary superposition of the basis states, $\ket{\pm}$. Without loss of generality, let us try to find a way to measure in the basis $\{\ket{\phi},\ket{\overline{\phi}}\}$ where $\ket{\phi}=c\ket{+}+d\ket{-}$ and $\ket{\overline{\phi}}=d^*\ket{+}-c^*\ket{-}$. By using the following rotation, we can rotate the state to the $\ket{\pm}$ basis,
\begin{align}
  U=\left(\begin{matrix}
    c^* & d^* \\
    d & -c \\
  \end{matrix}\right).
\end{align}
The result of the rotation is $U\ket{\phi}=\ket{+}$ and $U\ket{\overline{\phi}}=\ket{-}$. $\ket{+}$ and $\ket{-}$ are standard \schrodinger cat states, albeit with an additional phase, and can be distinguished using phase rotations and parity measurements. 

To implement the rotation $U$ one can use the approach of \citet{Ralph2003} in which $U$ is decomposed into Euler rotations which can be implemented nearly deterministically. The rotation above can be decomposed into
\begin{align}
  U(\theta,\lambda,\sigma)=iR_z(q)\hat{H}R_z(r)\hat{H}R_z(s)\label{rotationdecomposition},
\end{align}
where $\hat{H}$ is the Hadamard matrix, and $R_z(2\theta)=\one\cos\theta-i\sigma_z\sin\theta$ is the phase rotation gate \cite{Nielsen2000}. The phase rotation gate can be accomplished using a displacement operation and teleportation and the Hadamard matrix can be accomplished using a teleportation-like setup \cite{Ralph2003}. The values of $q$, $r$, and $s$ in \eqref{rotationdecomposition} are given by
\begin{align}
  q = \frac{\pi}{2}+\xi+\eta,\,\,\,
  r = 2\theta,\,\,\,
  s = \frac{\pi}{2}+\xi-\eta,
\end{align}
where $\xi=\arg c$, $\eta=\arg d$, and $\tan\theta=|d/c|$. It is easy to see that this approach, involving five non-deterministic operations will surely cause a reduction in the overall secret key fraction. Regardless, it is possible that an improved measurement approach may alleviate this problem.

We have shown a novel approach to implementing a device-independent quantum key distribution in which phase-entangled coherent states are used as the information carriers. We show how a judicious choice of initial states relative to the overall transmission rate of the channel can, in principle, lead to improved key generation rates relative to depolarizing channels in device-independent quantum key distribution systems based on polarization entanglement. We have shown that for QBERs exceeding $\sim 6.7\%$, phase-entangled states can be superior to polarization-entangled states, and we have given the explicit form of the states and measurements which achieve this result. We reiterate that apart from being secure against collective attacks and, in some instances, coherent attacks, our system is also flexible to time-varying loss in the quantum channel, for example, in ground-to-satellite schemes. This means, for example, that if the transmission loss suddenly decreases, the input states and measurements may be adapted to optimally make use of the updated channel. Finally, we note that our implementation does not require the eavesdropper to be restricted to Gaussian measurements.  


%
\end{document}